\documentstyle[11pt,newpasp,twoside]{article}
\def\edcomment#1{\iffalse\marginpar{\raggedright\sl#1\/}\else\relax\fi}
\reversemarginpar

\begin{document}
\title{A Flaring Megamaser in Mrk~348 }
 \author{Peck, A.~B., Falcke, H., Henkel, C., Menten, K.~M., Hagiwara, Y.}
\affil{MPIfR, Auf dem H\"ugel 69, D-53121 Bonn, Germany}
\author{Gallimore, J.~F. }
\affil{Bucknell Univ., Lewisburg, PA 17837 }
\author{Ulvestad, J.~S. }
\affil{ NRAO, P.O. Box 0, Socorro, NM 87801}

\begin{abstract}
We report new observations of the H$_2$O megamaser in the Seyfert 2
galaxy Mrk~348.  Following our initial detection in 2000 March using
the Effelsberg 100 m telescope, re-analysis of previous data on this
source indicates that the maser was present but only marginally
detectable in late 1997.  Monitoring through late 2000 shows that the
maser has again decreased to its original level.  The H$_2$O line is
redshifted by $\sim$130 km s$^{-1}$ with respect to the systemic
velocity, is extremely broad, with a FWHM of 130 km s$^{-1}$, and has
no detectable high velocity components within 1500 km s$^{-1}$ on
either side of the strong line.  Followup VLBA observations show that
the maser emission emanates entirely from a region $\le$0.25 pc in
extent, toward the base of the radio jet.

\end{abstract}
\section{Introduction}

Mrk~348 (NGC 262) is a Seyfert 2 galaxy at a redshift of 0.01503
(Huchra \mbox{\it et al.} 1999).  The galaxy is classified as an S0
with a low inclination, and exhibits a large H\kern0.1em{\sc i} halo
which may have been produced by an interaction with the companion
galaxy NGC 266 (e.g. Simkin \mbox{\it et al.} 1987).  VLBA images
(Ulvestad \mbox{\it et al.} 1999) reveal a small-scale double
continuum source, the axis of which is aligned with the optical
([OIII], Capetti \mbox{\it et al.} 1996) emission.  Astrometry
indicates that the optical and VLBI cores are nearly coincident.
Apparent subrelativistic expansion of the inner two jet components has
been detected by Ulvestad \mbox{\it et al.} (1999).  Ground-based
observations (Simpson \mbox{\it et al.} 1996) show evidence of a dust
lane crossing the nucleus and an ionization cone.  Attempts to detect
the expected obscuring torus at radio wavelengths
(e.g. H\kern0.1em{\sc i}, Gallimore \mbox{\it et al} 1999; CO,
Taniguchi \mbox{\it et al.} 1990; free-free absorption, Barvainis \&
Lonsdale 1998) have not been successful.

The compact radio source in Mrk~348 is unique among Seyferts in that
it is very bright and extremely variable.  The observations presented
here were made during a local minimum, when the total continuum flux
density at 22 GHz was $\sim$0.6 Jy.

\begin{figure}
\plotfiddle{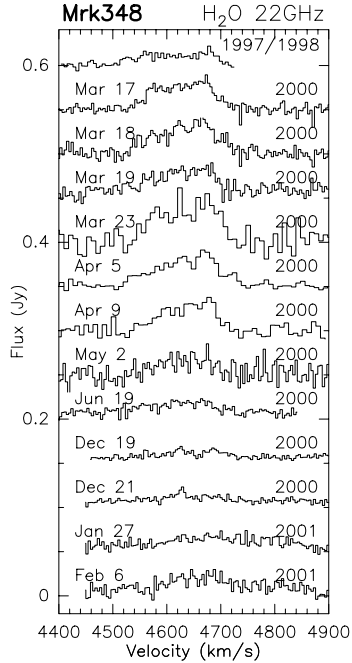}{8cm}{-90}{45}{45}{-80}{260}
\caption{Single dish profiles from Effelsberg 100m telescope.  The
peak flux in the line was $\sim$40 mJy on April 9, but decreased to 18
mJy by June 19.}
\end{figure}

\section{Observations}

The initial detection of the flaring maser in Mrk~348 using the
Effelsberg 100m telescope took place in 2000 March.  Re-analysis of
previous data on this source (top profile, Fig.~1) indicates that the
maser was also present but only marginally detectable in late
1997.  Monitoring through June 2000 showed
that the maser again decreased to its original level within 2 months,
as shown in Fig.~1.  The June 19 profile indicates a peak flux of
$\sim$18 mJy.  Resumption of the monitoring program in December 2000
showed little change in the line flux, though February 2001
observations yield a slight strengthening of the line.  We will
continue to monitor this source and additional follow-up observations
will be attempted if it is seen to flare again.

The H$_2$O maser line in Mrk~348 is extremely broad, with a FWHM of
$\sim$130 km s$^{-1}$, though in many of the monitoring epochs the
emission appears to consist of 2 lines which can be fit by Gaussian
functions with FWHM of $\sim$60 km s$^{-1}$ and $\sim$100 km s$^{-1}$
respectively, separated by $\sim$70 km s$^{-1}$.  There are no
detectable high velocity components within 1500 km s$^{-1}$ on either
side of the strong emission line.

The VLBA observations took place on 2000 June 10. March Effelsberg
observations indicated that the line was too broad to fit in a single
16 MHz IF (FWZP$>$250 km s$^{-1}$), so 2 IFs of 16 MHz each were used,
overlapped by 5 MHz.  Following calibration, the overlapping channels
were removed and the 2 IFs were added together to yield a single cube
of 174 channels covering 23 MHz.  Line profiles of the resulting cube
are shown in Fig.~2, superimposed on a continuum map made from 20
line-free channels. The maser emission is clearly seen to lie along
the line of sight to the jet, rather than the core which is thought to
lie toward the southern end of the continuum source.  The Gaussian fit
to the line shown in the first profile has an amplitude of 16$\pm$2
mJy and an integrated flux of 2.4$\pm$0.3 Jy/beam/km s$^{-1}$,
indicating that all of the flux measured in the Effelsberg 19 June
observation has been recovered.  The FWHM is 142$\pm$9 km s$^{-1}$
centered on V$_{\rm LSR}$=4640$\pm$2 km s$^{-1}$, consistent with the
single-dish measurements and redshifted by 131 km s$^{-1}$ with
respect to the systemic velocity.  A tentative 2 component fit to the
data yields a narrower line at 4683 km s$^{-1}$ with FWHM $\sim$60 km
s$^{-1}$ and amplitude $\sim$8 mJy, and a broader line at 4620 km
s$^{-1}$ with FWHM $\sim$100 km s$^{-1}$ and amplitude $\sim$12 mJy,
again consistent with our single-dish measurements.  No maser emission
is seen toward any other region of the radio source, only toward the
jet, and here the maser emission is unresolved at our angular
resolution of 0.42$\times$0.76 mas.  This corresponds to a linear size
of less than 0.25 pc (assuming H$_0$=75 km s$^{-1}$Mpc$^{-1}$).

\begin{figure}
\plotfiddle{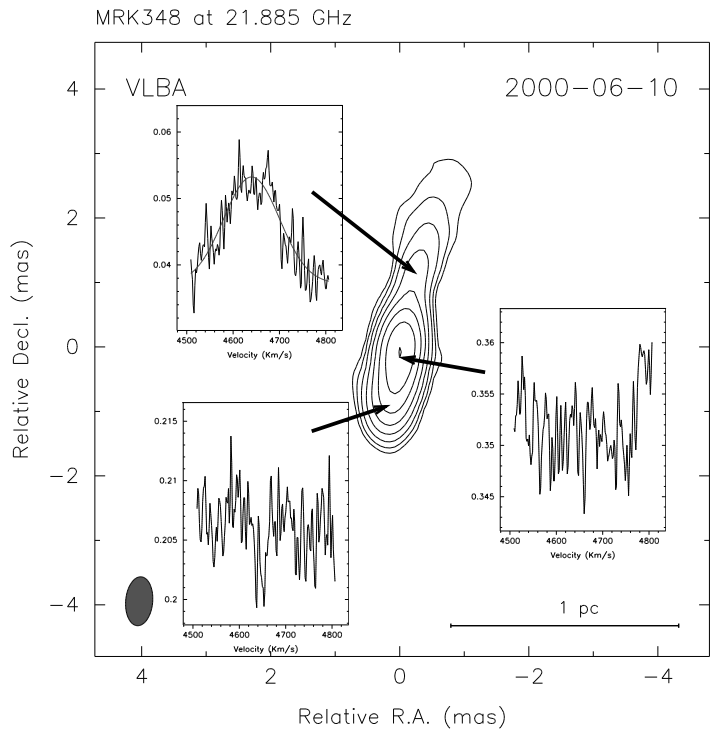}{9.5cm}{0}{50}{50}{-170}{-65}
\caption{Line profiles from VLBI data.  The continuum map is naturally
weighted.  The lowest contour is 3 mJy and the peak is 352 mJy.  The
RMS noise in the continuum map is $<$1 mJy/beam, and in the line
profiles is $\sim$4 mJy/beam/channel.}
\end{figure}

\section{Conclusions}

During early 2000, the H$_2$O emission toward Mrk~348 showed a
dramatic intensity increase which coincided with a significant
increase in the flux of the nuclear radio continuum source.  The
unusual line profile leads us to suspect that this source, and
possibly NGC~1052, might belong to a class of megamaser galaxies in
which the amplified emission is the result of an interaction between
the radio jet and an encroaching molecular cloud, rather than
occurring in a circumnuclear disk.  Analysis of our recent VLBA
observations indicates that the emission does indeed arise along the
line of sight to the jet in Mrk~348 (see Fig.~2), confirming this
prediction.  The very high linewidth occurring on such small spatial
scales indicates that the H$_2$O emission arises from a shocked region
at the interface between the energetic jet material and the molecular
gas in the cloud where the jet is boring through.  This hypothesis is
supported by the spectral evolution of the continuum source
(Brunthaler, priv. comm.), which showed an inverted radio spectrum
with a peak at 22 GHz, later shifting to lower frequencies. By analogy
to IIIZw2 (Brunthaler \mbox{\it et al.} 2000) this would indicate the
formation and evolution of very compact hotspots propagating through a
dense medium.  In this scenario, the recent high frequency radio
continuum flare and the northward movement of the brightest continuum
component are attributable to the impact between the jet and the
molecular cloud.  The very close temporal correlation between the
flaring activity in the maser emission and the continuum flare further
suggest that the masing region and the continuum hotspot are nearly
coincident and may be different manifestations of the same dynamical
events.

\acknowledgements

We would like to thank Barry Clark and the staff at NRAO for promptly
scheduling and correlating the VLBA observations as a Target of
Opportunity experiment.  The National Radio Astronomy Observatory is a
facility of the National Science Foundation operated under a
cooperative agreement by Associated Universities, Inc.


\vspace{-0.25cm}
\begin{references}

{\small
\reference Barvainis, R. \& Lonsdale, C. 1998, \aj, 115, 885
\reference Brunthaler, A., Falcke, H., Bower, G.~C., Aller, M.~F., Aller, H.~D.,Ter\"asranta, H., Lobanov, A.~P., Krichbaum, T.~P. \& Patnaik, A.~R. 2000, \aap, 357, L45
\reference Capetti, A., Axon, D.~J., Macchetto, F., Sparks, W.~B. \& Boksenberg, A. 1996, \apj, 469, 554
\reference Falcke, H. Henkel, C., Peck, A.~B., Hagiwara, Y., Prieto, M.~A. \&
 Gallimore, J.~F. 2000, \aap, 358, L17 
\reference Gallimore, J.~F., Baum, S.~A., O'Dea C.~P., Pedlar, A. \& Brinks, E. 1999, \apj, 524, 684
\reference Huchra, J.~P., Vogeley, M.~S. \& Geller, M.~J. 1999, \apjs, 121, 287
\reference Simkin, S.~M, Su, H.-J., van Gorkom, J. \& Hibbard, J.  1987, Science, 235, 1367
\reference Simpson, C., Mulchaey, J.~S., Wilson, A.~S., Ward, M.~J. \& Alonso-Herrero, A. 1996, \apj, 457, L19
\reference Taniguchi, Y., Kameya, O., Nakai, N. \& Kawara, N. 1990, \apj, 358, 132 
\reference Ulvestad, J.~S., Wrobel, J.~M., Roy, A.~L., Wilson, A.~S., Falcke, H \& Krichbaum, T.~P. 1999, \apj, 517, L81
}
\end{references}
\end{document}